\documentclass[8.5pt,twoside,twocolumn]{article}
\usepackage[version=3]{mhchem}
\usepackage[super,sort&compress,comma]{natbib} 
\usepackage{mhchem}
\usepackage{times,mathptmx,amssymb}
\usepackage{sectsty}
\usepackage{balance} 

\usepackage{graphicx} 
\usepackage{lastpage}
\usepackage[format=plain,justification=raggedright,singlelinecheck=false,font=small,labelfont=bf,labelsep=space]{caption} 
\usepackage{fancyhdr}
\begin{document}

\thispagestyle{plain}
\fancypagestyle{plain}{
\renewcommand{\headrulewidth}{1pt}}
\renewcommand{\thefootnote}{\fnsymbol{footnote}}
\renewcommand\footnoterule{\vspace*{1pt}%
\hrule width 3.4in height 0.4pt \vspace*{5pt}} 
\setcounter{secnumdepth}{5}

\makeatletter 
\def\subsubsection{\@startsection{subsubsection}{3}{10pt}{-1.25ex plus -1ex minus -.1ex}{0ex plus 0ex}{\normalsize\bf}} 
\def\paragraph{\@startsection{paragraph}{4}{10pt}{-1.25ex plus -1ex minus -.1ex}{0ex plus 0ex}{\normalsize\textit}} 
\renewcommand\@biblabel[1]{#1}            
\renewcommand\@makefntext[1]%
{\noindent\makebox[0pt][r]{\@thefnmark\,}#1}
\makeatother 
\renewcommand{\figurename}{\small{Fig.}~}
\sectionfont{\large}
\subsectionfont{\normalsize} 

\fancyfoot{}
\fancyhead{}
\renewcommand{\headrulewidth}{1pt} 
\renewcommand{\footrulewidth}{1pt}
\setlength{\arrayrulewidth}{1pt}
\setlength{\columnsep}{6.5mm}
\setlength\bibsep{1pt}

\twocolumn[
  \begin{@twocolumnfalse}
\noindent\LARGE{\textbf{Detached topological charge on capillary bridges$^\dag$}}
\vspace{0.6cm}

\noindent\large{\textbf{Verena Schmid,\textit{$^{a}$}  and
Axel Voigt$^{\ast}$\textit{$^{a,b}$}}}\vspace{0.5cm}

\noindent\textit{\small{\textbf{Received Xth XXXXXXXXXX 20XX, Accepted Xth XXXXXXXXX 20XX\newline
First published on the web Xth XXXXXXXXXX 200X}}}

\noindent \textbf{\small{DOI: 10.1039/b000000x}}
\vspace{0.6cm}

\noindent \normalsize{We numerically investigate crystalline order on negative Gaussian curvature capillary bridges. In agreement with the experimental results in [W. Irvine \textit{et al., Nature, Pleats in crystals on curved surfaces}, 2010, \textbf{468}, 947]} we observe for decreasing integrated Gaussian curvature a sequence of transitions, from no defects to isolated dislocations, pleats, scars and isolated sevenfold disclinations. We especially focus on the dependency of the detached topological charge on the integrated Gaussian curvature, for which we observe, again in agreement with the experimental results, no net disclination for an integrated curvature down to -10, and a linear behaviour from there on until the disclinations match the integrated curvature of -12. The results are obtained using a phase field crystal approach on catenoid-like surfaces and are highly sensitive to the initialization.
\vspace{0.5cm}
 \end{@twocolumnfalse}
  ]

\footnotetext{\dag~Electronic Supplementary Information (ESI) available: [Animation of the evolution of defect motifs on a surface with increasingly negative integrated Gaussian curvature.]. See DOI: 10.1039/b000000x/}


\footnotetext{\textit{$^{a}$~Institute of Scientific Computing, TU Dresden, 01062 Dresden, Germany}}
\footnotetext{\textit{$^{b}$~Centre of Advanced Modeling and Simulation, TU Dresden, 01062 Dresden, Germany; E-mail: axel.voigt@tu-dresden.de}}




Geometric and topological constraints provide a route to control the self-assembly of simple building blocks of soft materials. This has e.g. been addressed in the themed collection on {\em The geometry and topology of soft materials} \cite{SM_34_2013_8075}. One of the key concepts behind these phenomena is the notion of geometric frustration. When soft materials assemble in the presence of geometric and topological constraints, the regular order favored by local interactions is frustrated, which leads to defect structures in the ground state. These defects can be chemically functionalized and provide the key to self-assembly into complex hierarchical structures with emergent novel macroscopic properties. An understanding of the interplay of geometry, topology and defects is therefore of utmost importance, if this route to design novel materials will be explored. 

\begin{figure}[h]
\centering
  \includegraphics[height=2.33cm]{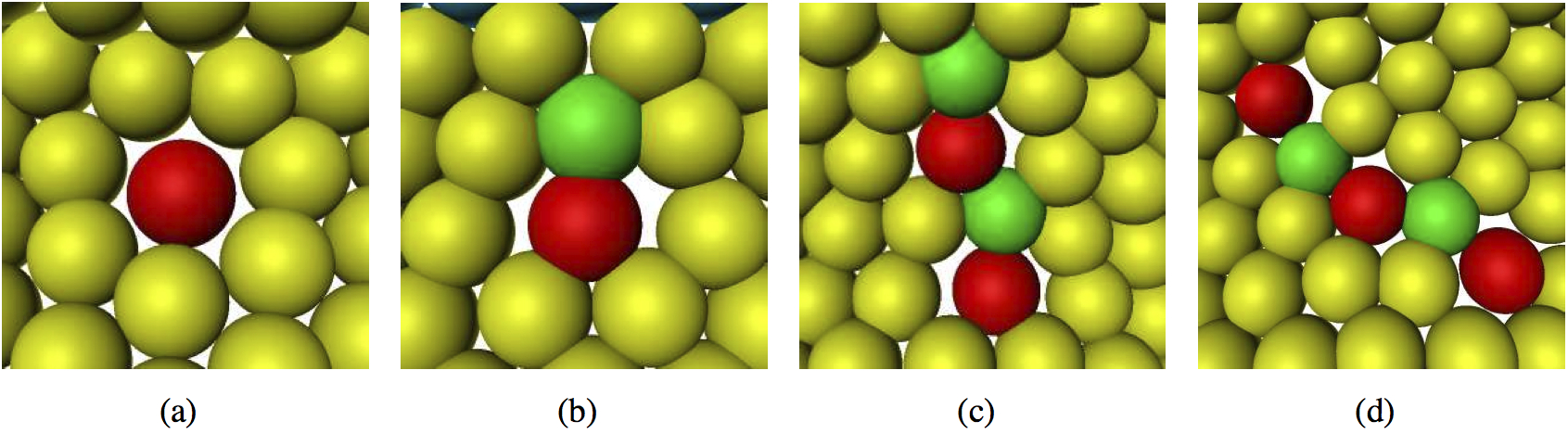}
  \caption{Defect types: (a) sevenfold disclination, (b) dislocation, (c) pleat consisting of two dislocations, (d) scar with topological charge of $-1$. The colour coding corresponds to the number of neighbors: 
 green (5), yellow (6) and red (7), within Fig. \ref{fig4} we will find dark blue (3) and light blue (4).}.
  \label{fig1}
\end{figure}

Various theoretical results are known. A classic theorem of Euler, e.g. shows for a triangulation of the surface in which nearest neighbors are connected, which corresponds to the order resulting from local interactions, that $\sum_i (6 - i) v_i = 6\xi$, with $v_i$ as the number of vertices with $i$ nearest neighbors and $\xi$ as the Euler characteristic of the surface. Thus for surfaces with the topology of a sphere $(\xi =2)$, besides the expected triangular lattice with sixfold coordination, which would give the optimal packing in a plane, there must be at least 12 fivefold disclinations present. However, with each disclination an extra energy is associated (relative to a perfect triangular lattice in flat space) which grows proportional to $R^2$, with $R$ as the radius of the sphere. For a fixed lattice constant $a$ we have $N \sim (R/a)^2$, with $N$ the number of particles. Thus for large $N$ mechanisms are expected which reduce this extra energy by changing the ground-state configuration. In cases where large surface tension limits significant buckling the energy is reduced by introducing dislocations and grain-boundary scars. Realizations of water droplets in oil, which are coated with colloidal particles \cite{Bauschetal_Science_2003} show that these excess dislocations grow linearly with the system size. The same arguments hold for capillary bridges with the topology of a cylinder $(\xi = 0)$. In this case, there is no topological need for disclinations and all defects are introduced to relieve the strain induced by the curvature. Experiments for colloidal particles on oil-glycerol capillary bridges \cite{Irvineetal_Nature_2010} show a more complex behaviour than on spheres. No net topological charge is found on the surface for an integrated Gaussian curvature down to $\Omega = -10$, with $\Omega = 3 / \pi \int_\Gamma G \; d \Gamma$ and Gaussian curvature $G$ . For $\Omega$ beyond the threshold of $-10$, disclinations rapidly fill the surface until 12 disclinations approximately match the integrated curvature of $-12$. A sequence of transitions is observed, from no defects to isolated dislocations, which organize into pleats, and finally form scars and isolated sevenfold disclinations, see Fig. \ref{fig1} for an explanation of the various defect types. 

While the linear increase of excess dislocations and the formation of scars on a sphere is well understood and reproduced quantitatively by various theoretical approaches and simulation techniques, see e.g. \cite{Travesset_PRE_2005,Schneideretal_EPL_2005,Walesetal_PRB_2006,Vitellietal_PNAS_2007,Backofenetal_PRE_2010}, and the review \cite{Bowicketal_AP_2009}, the behaviour on capillary bridges is less understood. In \cite{Bowicketal_EPL_2011} the problem is addressed by mapping the miscroscopic interacting particle problem to the problem of discrete interacting defects in a continuum elastic background. The critical value
for the transition between configurations with and without disclinations or scars is deduced for catenoids. Three different ways are considered. In the first approach the critical value corresponds to the catenoid for which the integrated Gaussian curvature of the geodesic disc matches the curvature, which would be screened by a sevenfold disclination. Using the definition in \cite{Irvineetal_Nature_2010}, in which the integrated Gaussian curvature is defined in units of disclinations as above, the obtained critical waist size $c^* = 0.85$ corresponds to $\Omega^* = - 6.3$. The second approach uses an effective screened disclination charge, for which $c^* = 0.87$, corresponding to $\Omega^* = -6.0$. The third approach uses an energetic argument. The transition point for the emergence of a disclination in the interior is found to be $c^* = 0.8$, corresponding to $\Omega^* = -7.2$. All values are significantly larger than the experimentally obtained $\Omega^* = -10$. The same approach is used in \cite{Irvineetal_SM_2012} and validated against a numerical energy minimization of a discrete spring model. The results are in qualitative agreement with \cite{Irvineetal_Nature_2010} but no quantitative estimate for $\Omega^*$ is given. Simulation results for electrostatically charged particles are obtained for a wide range of system size, curvature and interaction potentials, including Yukawa, Coulomb and Lennard-Jones in \cite{Kusumaatmajaetal_PRL_2013,Benditoetal_PRE_2013}. The results qualitatively agree with the continuum theory and the experimental data. Depending on $\Omega$ dislocations, pleats, scars and sevenfold disclinations are found, independent of the used potential. The supplemental material in \cite{Kusumaatmajaetal_PRL_2013} also provides tabulated numbers of positive and negative topological charges and a critical waist size $c^*$, which here correspond to the transition to isolated sevenfold disclinations. It seems to be only weakly dependent on the number of particles, but sensitive to the considered potential. For the Coulomb potential, a critical waist size $c^* = 0.55$ is found, corresponding to $\Omega^* = -10.01$. However, the tabulated charges indicate the presence of charged defects as scars already above that value, in disagreement with the experimental results in \cite{Irvineetal_Nature_2010}. The data also strongly depend on the considered constant mean curvature (Delauny) surface. Within the experiments, the capillary bridge is obtained through a sequence of Delaunay surfaces (unduloids, catenoids and nodoids) each with different geometrical parameters, such as aspect ratio, mean curvature and maximal Gaussian curvature. Such different Delaunay surfaces are considered in \cite{Benditoetal_PRE_2013}. However, quantitative results for $\Omega^*$ are not given.

Here, instead of a coarse grained elasticity problem for the defects or a discrete particle simulation for the position of $N$ particles, we consider a continuous optimization problem for a number density, with $N$ maxima representing the $N$ particles. The idea is to find a free energy, which is minimized for a configuration, which corresponds to the minimum of the original problem. We consider the Swift-Hohenberg free energy on the surface $\Gamma$ with $\nabla_\Gamma$ and $\Delta_\Gamma$ the corresponding surface gradient and surface Laplacian. We define 
\begin{eqnarray}
  \label{eq:free_energy_surface}
  {\cal F}^\Gamma[\rho] = \int_\Gamma - |\nabla_\Gamma \rho|^2 + \frac{1}{2} |\Delta_\Gamma \rho|^2 + f(\rho) \; \mathrm{d}\Gamma.
\end{eqnarray}
Thereby $\rho$ denotes a number density of the particles and $f(\rho) = \frac{1}{2}(1 - \epsilon)\rho^2 + \frac{1}{4} \rho^4$ can have a double-well structure, depending on the parameter $\epsilon$. Within various parameter regimes, the minimizers of this energy are periodic solutions with a hexagonal structure in flat space, which are geometrically frustrated in the considered setting on $\Gamma$. We here have an interplay between the tendency to form periodic solutions and the geometric frustration which prevent such a periodicity. We consider the $H^{-1}$ gradient flow of this energy, which can be viewed as an extension of the phase-field-crystal (PFC) model, introduced in \cite{Elderetal_PRL_2002} to surfaces. From 
\begin{eqnarray}
\partial_t \rho &=& \Delta_\Gamma \frac{\delta {\cal F}^\Gamma[\rho]}{\delta \rho}
\end{eqnarray}
we obtain the following system of surface partial differenial equations
\begin{eqnarray}
  \label{eq_1}
  \partial_t \rho &=& \Delta_\Gamma \mu \\
  \label{eq_2}
  \mu &=& 2 \nu + \Delta_\Gamma \nu + f^\prime(\rho) \\
  \label{eq_3}
  \nu &=& \Delta_\Gamma \rho
\end{eqnarray}
In \cite{Backofenetal_PRE_2010} this phenomenological approach is validated for the Thomson problem \cite{Thomson_PM_1904} by computing minimal energy configurations for various numbers of $N$ and comparing the resulting configurations and Coulomb energies with known results for $N \in [12, 2790]$. However, the approach is also quantitatively related to discrete particle simulations. In \cite{Backofenetal_MMS_2011} the equations are derived from a discrete particle setting via classical dynamic density functional theory, following the derivation in \cite{Teeffelenetal_PRE_2009} for flat space. Numerical results on a sphere have shown, that the obtained minimal energy configurations are insensitive to the specific underlying interparticle potential. These results are in agreement with \cite{Kusumaatmajaetal_PRL_2013}, where five different potentials are considered and qualitatively the same defect motifs are found. Both results indicate that the geometric frustration is much stronger than the influence of the interparticle potential. We therefore stick with the phenomenological model above. Note that with this model the number of particles $N$ is not fixed throughout evolution and can therefore not set to a distinct value.

Different numerical approaches have been considered in \cite{Backofenetal_MMS_2011}, we here adapt a parametric finite element setting, which is realized within the simulation toolbox AMDiS \cite{Veyetal_CVS_2007}. The approach is thereby based on the stable finite element discretization for the PFC model in \cite{Backofenetal_PML_2007} and is described in detail in \cite{Backofenetal_MMS_2011}. The key idea is to use the surface operators on the discrete surface which consists of triangles $T$. To do the integration on these triangles a parameterization $F_{T}: \hat{T} \to T$ is used, with $\hat{T}$ the standard element in $\mathbb{R}^2$. These allow to transform all integrations onto the standard element using the finite element basis defined also in $\mathbb R^{2}$. The parameterization $F_T$ is given by the coordinates of the surface mesh elements and provides the only difference between solving equations on surfaces and on planar domains. For a surface we have to allow $F_T: \mathbb{R}^2 \to \mathbb{R}^{3}$, whereas for a planar domain $F_T: \mathbb{R}^2 \to \mathbb{R}^2$. With this tiny modification any code to solve partial differential equations on Cartesian grids can be used to solve the same problem on a surface, providing a surface triangulation is given. It is essentially this modification which induces the geometric frustration to the problem.

\begin{figure}[h]
\centering
  \includegraphics[height=2.4cm]{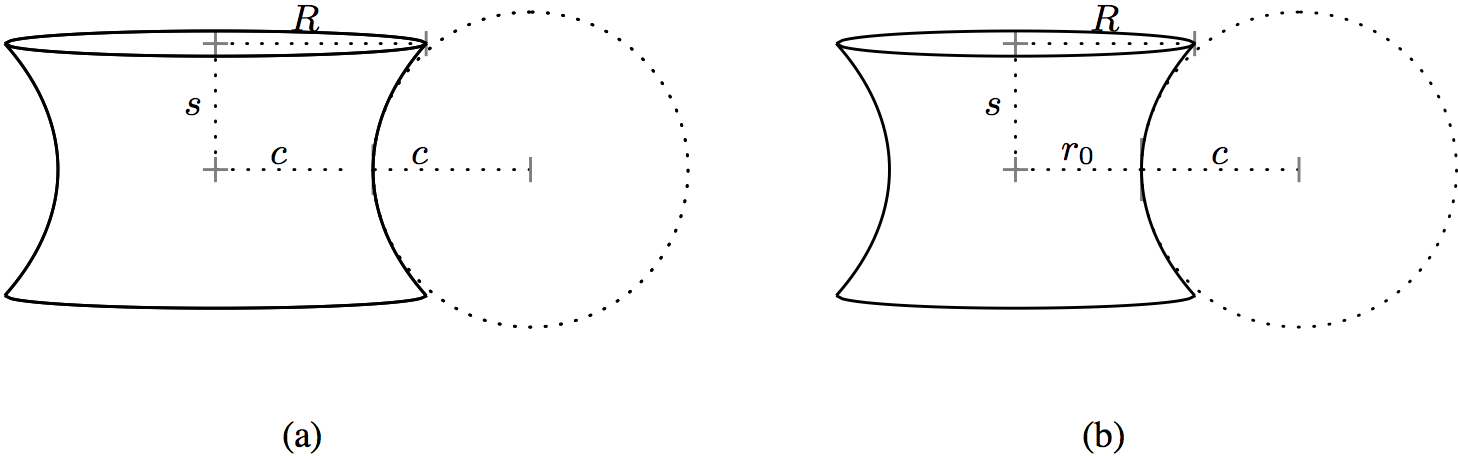}
  \caption{Schematic figure of a catenoid (a) and the considered approximation (b), for which $r_0 = s = \frac{4}{5}c$.}.
  \label{fig2}
\end{figure}

Our surface is given as an approximation of a catenoid and obtained through a rotation around the $x_3$-axis. The parametric equation of the surface is given by
\begin{eqnarray}
\Phi(t,\phi) = \left( \begin{array}{c} (c \cosh \frac{t}{c} + (r_0 -c)) \cos \phi \\ (c \cosh \frac{t}{c} + (r_0 -c)) \sin \phi \\ t \end{array} \right)
\end{eqnarray}
Only for $r_0 = c$ the approximation is equal to a catenoid, see Fig. \ref{fig2}. In the other cases, it does not have a constant mean curvature and is especially no minimal surface, but provides a satisfying approximation for the considered experiments in \cite{Irvineetal_Nature_2010}, see suppl. information. 

The surface area is computed as $A = 2 \pi c ( c \sinh \frac{s}{c} \cosh \frac{s}{c} + s + 2(r_0 -c) \sinh \frac{s}{c})$, the Gaussian curvature follows as $G(t) = - 1 / (c(c \cosh \frac{t}{c} + r_0 -c) \cosh^3 \frac{t}{c})$ and $\Omega = -12 \tanh \frac{s}{c}$. Using this approximation, we can scale the surface area by varying the waist radius $r_0$, without changing the height $s$, the vertical curvature radius at the waist $c$ and the integrated Gaussian curvature $\Omega$. The approach allows to change the surface to scatter $\Omega \in (-12,0]$ by keeping the surface area $A$ and the outer radius $R$ fixed.

The surface is discretized using triangular elements with a sufficient mesh size $h \sim a/10$. We use the parameters $\epsilon = 0.4$ and the average particle density as $\rho_0 = -0.3$, which correspond to a point in the two-dimensional phase diagram within the hexagonal region. At the boundary we describe Dirichlet conditions using an one-mode approximation, see \cite{Elderetal_PRE_2004}.

\begin{figure}[h]
\centering
  \includegraphics[height=1.3cm]{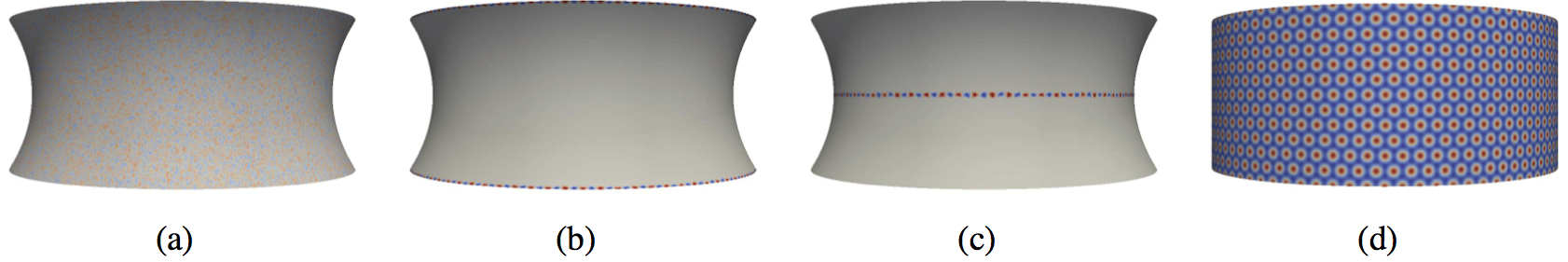}
  \caption{Types of initialization: (a) random initialization with $\rho = \rho_0 + \eta$, with white noise $\eta$, (b) initialization at the boundary, with the one-mode approximation specified at the boundary, (c) initialization at the waist, with the one-mode approximation specified at the waist and (d) a cylinder with the one-mode approximation on the surface. In all cases the particle density $\rho$ is shown.}.
  \label{fig3}
\end{figure}

We consider four different initializations, see Fig. \ref{fig3}. We either specify different initial conditions or as in \cite{Irvineetal_Nature_2010} start with a cylinder and subsequently change the surface to decrease $\Omega$. Within this last approach we use a sequence of 42 geometries and compute the steady state on each, with the one from the previous surface as initial condition. An animation is provided in the ESI \dag.

The simulation results are postprocessed in the following way. We identify the maxima of the computed particle density $\rho$, interprete them as particle positions and define their neighborship based on their two-dimensional Voronoi regions. These data are used to evaluate the number of dislocations, scars, pleats, five- and sevenfold disclinations, as well as the number of three- and fivefold disclinations on the boundary. Within colliding defect structures, clusters are seperated preferring dislocations and pleats over disclinations and scars, larger defects over smaller defects and oriented defects over unoriented defects. The visualization is done using the software {\em Ovito} \cite{Stukowski_MSMSE_2010}. 

\begin{figure*}[t]
  {\centering
  \includegraphics[height=9.3cm]{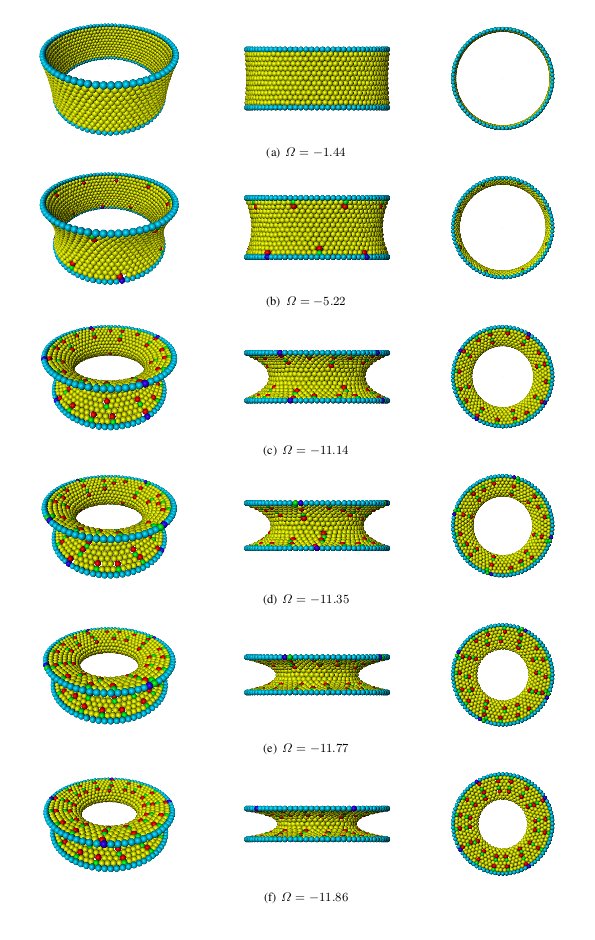}
 \includegraphics[height=9.3cm]{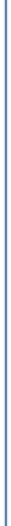}
  \includegraphics[height=9.3cm]{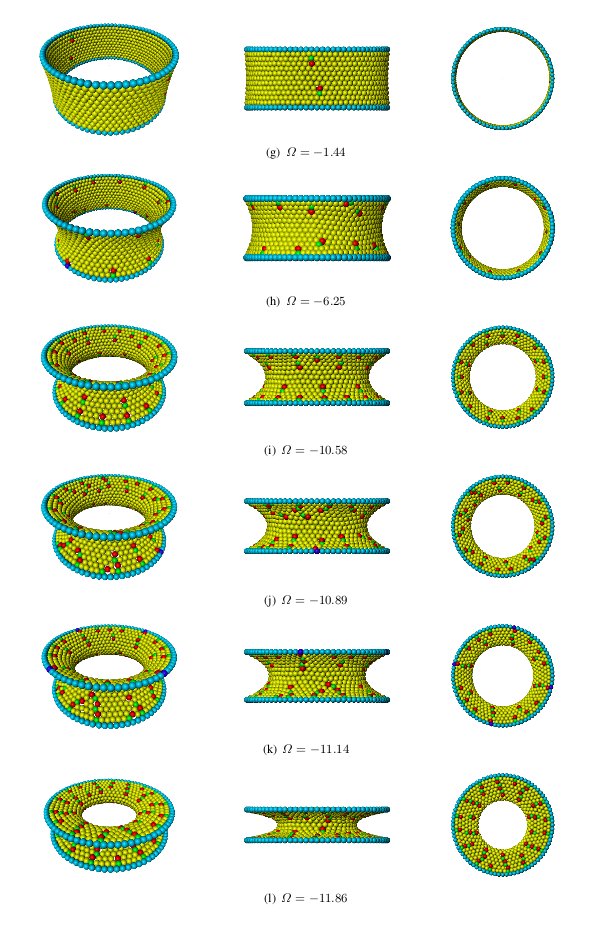}
 \includegraphics[height=9.3cm]{line.jpg}
  \includegraphics[height=9.3cm]{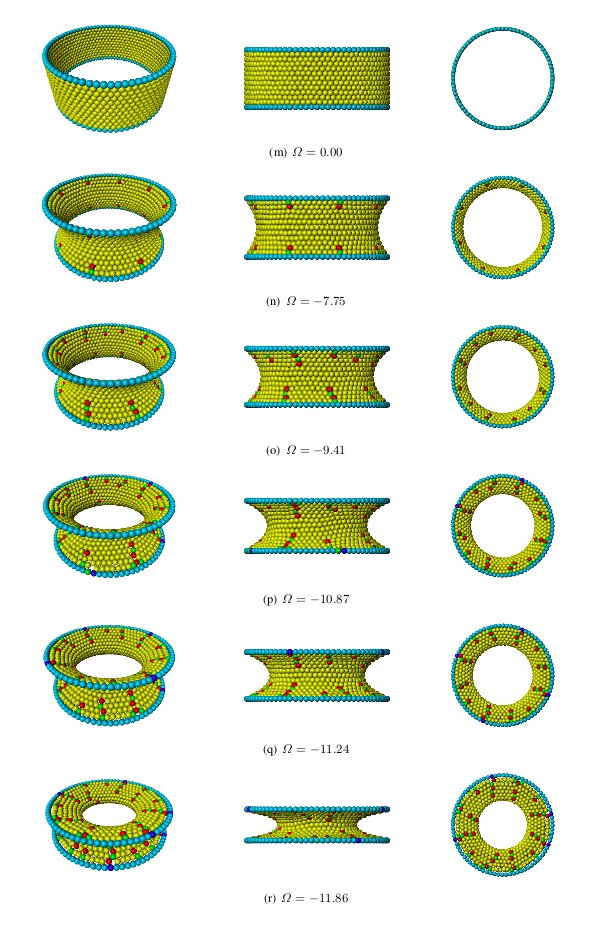}} \\
  (1) initialization from boundary \hspace*{1.5cm} (2) initialization from waist \hspace*{2.2cm} (3) evolving surface
  \caption{Final configuration for six selected values of $\Omega$ for three different initializations. The configurations are selected to show characteristic features: (a) shows the largest and (f) the smallest considered value for $\Omega$. (b) shows the first appearance of defects, which might still be connected to the boundary, or an intermediate value if defects are already present for the initial value of $\Omega$, (c) and (d) show the first appearance of pleats or an intermediate configuration, and (d) and (e) the first appearance of scars and sevenfold disclinations. For the evolving surface an animation through all considered 42 surfaces is provided in the ESI \dag.}
  \label{fig4}
\end{figure*}

Fig. \ref{fig4} shows selected results for three different initializations for decreasing $\Omega$. We thereby always show the results for largest and smallest $\Omega$, as well as four surfaces in between, which in most cases mark the onset or the disappearence of a defect type. For all initializations, also for the not shown random initialization, we observe the whole spectrum of defects: dislocations, pleats, scars, five- and sevenfold disclinations and the expected sequence of transitions to dislocations, pleats, scars and isolated sevenfold disclinations. Table \ref{tab1} summarizes their first appearance. 
  
\begin{table}[h]
\small
  \caption{The largest value of $\Omega$ for the appearance of different defect types depending on the initialization.}
  \label{tab1}
  \begin{tabular*}{0.48\textwidth}{@{\extracolsep{\fill}}lrrrr}
    \hline
                        & random & boundary & waist  & evolution \\
    \hline
    dislocations & -1.44    & -4.07      & -1.44   &  -7.75 \\
    pleats           & -4.07   &  -9.78      & -1.44   &  -9.41 \\
    scars            & -6.25   &   -11.35   & -10.89  &  -10.44 \\
     5-fold          & -9.27    & -11.35   & -11.77 & -10.87 \\
    7-fold          &  -11.53 & -11.35    & -11.14  & -10.87 \\
   \hline
  \end{tabular*}
\end{table}

The appearance and interactions of defects differ for different initializations. While for random initialization defects are already present for the largest value of $\Omega$, scars are mainly accompanied by fivefold disclinations next to the boundary and sevenfold disclinations appear at the waist, the configurations resulting from initialization at the boundary is characterized by migration of dislocations into the surface and the formation of a second row of dislocations, see Fig. \ref{fig4}, (d) in the left column, as well as an increase in size of the occuring pleats, see (e) in the left column. For the initialization from the waist, we observe only oriented dislocation for large $\Omega$, with the number increasing with decreasing $\Omega$ until the first defects occur also at the boundary, see (b) and (c) in the middle column. The appearance of pleats comes together with non-oriented dislocations. As the pleats grow, the dislocations become again oriented. For the evolving geometry, the surface remains defect free up to relatively low values for $\Omega$, first defects occur as oriented dislocation at the boundary, see (b) in the right column. Their number stays fixed until pleats are formed after further decreasing $\Omega$, see (c) in the right column. Here pleats do not grow but fall apart, forming an oriented dislocation in the interior and one at the boundary. Scars and sevenfold disclinations are formed at the same time, see (e) in the right column. In contrast with the size dependent exclusive appearence of disclinations or scars reported in \cite{Bowicketal_EPL_2011}, we found both defect types at the same time. 

For random initializations, scars already appear on surfaces with $\Omega = -6.25$. For the initialization at the boundary and the waist the configuration stays without detached topological charge down to $\Omega = -11.14$ and $\Omega = - 11.35$, respectively. All in quantitative disagreement with the experimental results in \cite{Irvineetal_Nature_2010}. Only the evolving geometry, which resembles the evolution in the experimental setting best, leads to the expected behavior, with no detached topological charge down to $-10.44$. 

\begin{figure*}[th]
  {\centering
  \includegraphics[height=4.7cm]{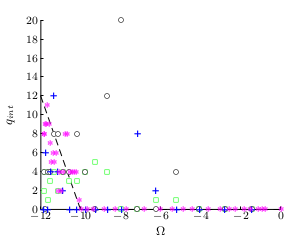}
  \includegraphics[height=4.7cm]{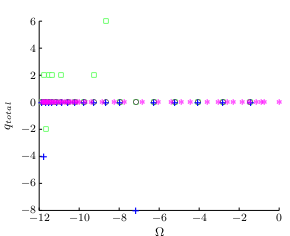}
  \includegraphics[height=4.7cm]{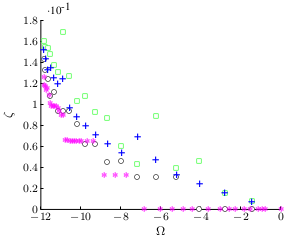}} 
  \caption{(left) detached topological charge $q_{int}$, (middle) total topological charge $q_{total}$, and (right) scaled number of defects $\zeta$ as function of $\Omega$, with green random initialization, black initialization from the boundary, blue initialization from the waist, and purple the evolving surface computation. The left figure shows in addition the experimentally observed  linear increase from 0 to 12 for $\Omega = -10$ to $\Omega = -12$ in \cite{Irvineetal_Nature_2010}.}
  \label{fig5}
\end{figure*}

To further highlight the quantitative agreement, we compute the detached topological charge $q_{int}$, as the difference of the number of particles in the interior with seven and five neighbors, the total topological charge $q_{total}$, which in addition considers particles at the boundary, which do not have four neighbors and thus should be zero for $\Omega \in (-12,0]$, and the number of defects $N_{defect}$, which we scale by the number of particles $N$ to obtain the ratio $\zeta = N_{defect} / N$. Fig. \ref{fig5} shows these characteristic quantities as a function of $\Omega$.

\begin{figure}[h]
  {\centering
  \includegraphics[height=3.3cm]{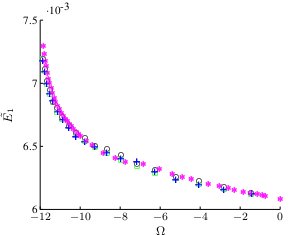}
  \includegraphics[height=3.3cm]{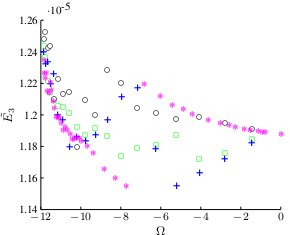}} 
  \caption{(left) scaled energy $E_1$, and (right) scaled energy $E_3$ as function of $\Omega$, with green random initialization, black initialization from the boundary, blue initialization from the waist, and purple the evolving surface computation.}
  \label{fig6}
\end{figure}

The detached topological charge for the evolving surface simulations show the expected behavior, with $q_{int} = 0$ down to $\Omega \approx -10$ and an approximately linear increase up to $q_{int} \approx 12$ for $\Omega \approx -12$. This is in quantitative agreement with the experimental results in \cite{Irvineetal_Nature_2010}. In agreement with the topological requirement the total charge $q_{total}$ remains zero up to the smallest considered value of $\Omega = - 11.86$ for all initializations, which is just a consistancy check of our postprocessing, and the number of defects growth for decreasing $\Omega$. For the evolving surface simulations the configurations stay defect free down to $\Omega = - 7.75$, where defects form suddenly. The number of defects stays almost constant afterwards until a second jump can be seen for $\Omega = -10.44$. Below that value more and more defects are introduced. This has a clear influence on the energy, which we compute as
\begin{eqnarray}
E_1 &=& \sum_{1 \leq i<j \leq N} \frac{1}{|p_i - p_j|}, \\ 
E_3 &=& \sum_{1 \leq i<j \leq N} \frac{1}{|p_i - p_j|^3} 
\end{eqnarray}
with $i$-th particle position $p_i$ and the distance measured in $\mathbb{R}^3$, which we again rescale to eliminate the dependence on $N$ to obtain $\tilde{E}_1 = E_1 / N^2$ and $\tilde{E}_3 = E_3 / N^2$. Fig. \ref{fig6} shows the computed energies as a function of $\Omega$. 

While the different initializations lead to almost identical results for long-range potential $\tilde{E}_1$, large differences can be seen for the short range potential $\tilde{E}_3$. The curve for the evolving surface resembles nicely the characteristic defects. The energy is increasing down to $\Omega = -7.75$ with a defect-free configuration. The sudden drop in the energy results from the occurance of defects, in the form of dislocations at the boundary. The energy again increases until $\Omega = - 10.44$. The small drop here corresponds again to a change in defect type, which leads to the occurance of detached topological charge.
The energy plots further indicate, that the configurations obtained with the evolving surface approach not always lead to the lowest energy. We can assume a highly complex energy landscape, in which all our observed configurations are presumably trapped in local minima. Instead of searching for global minima, as in \cite{Kusumaatmajaetal_PRL_2013}, we concentrate on resampling the experiments in \cite{Irvineetal_Nature_2010}, with presumable also only local minima configurations. With the evolving surface approach, which resamples the experimental setting best, quantitative agreement for all considered data $q_{int}$ and $\zeta$ can be achieved. The number of defects is thereby lower than in various of the putative global minima configurations reported in \cite{Kusumaatmajaetal_PRL_2013}.

The goal to chemically functionalize the defects to control self-assembly into supramolecular structures, requires not only the presence of a certain number and type of defects, but also their position and arrangement to be predictable. The configurations obtained with the evolving surface approach lead to highly symmetric arrangements, see Fig. \ref{fig3} (left column). The number of defects growth until 12 equally spaced oriented dislocations on the boundary are formed. These defects remain and grow into pleats, without changing their position. This highly symmetric arrangement even remains for smaller $\Omega$ after the splitting into scars and sevenfold disclinations and only for the lowest values of $\Omega$ the symmetry is lost. To identify this symmetric sequence of transitions as the most favorable path requires to compute the energy barriers and minimal energy paths, which is currently under investigation using the string method \cite{Renetal_PRB_2002} already applied to the phase field crystal model in \cite{Backofenetal_JPCM_2010,Backofenetal_EPJST_2014}. 
 
This research was supported by DFG within SPP 1296 Grant No. Vo899/7-3 as well as by EU within FP7 Grant No. 247504. 




\footnotesize{
\bibliography{SM} 
\bibliographystyle{rsc} 
}

\end{document}